\documentstyle[editedvolume]{crckapb} 

\begin{opening}
\title{Detecting cosmic rays of the highest energies\protect\\
       }
\subtitle{NATO Advanced Research Workshop, Oujda, Marocco, 21-23 March 2001}

\author{F. VANNUCCI}
\institute{LPNHE, Univ. Paris 7\\
           4 place Jussieu 75252 Paris, France}
\end{opening}

\runningtitle{DETECTING COSMIC RAYS}

\begin{document}

% The \begin{document} command comes after the \end{opening}
% command.

\section{Introduction}

The highest energies are defined, at a given time, by our
detector capabilities. Thus the highest energies will be 
the just measured or soon to be measured energies. To-day,
the frontier in cosmic rays reaches $10^{20}$ eV. This represents a
macroscopic energy, namely 16 J, transported by a microscopic object.
For charged particles, the highest energy events seem to necessitate 
some New Physics, beyond the present understanding of astrophysics and particle physics. 

Concerning neutrinos, apart from the low energy ones coming from the explosion of SN1987A, only neutrinos of terrestrial origin have been detected so far.
It is a challenge to detect neutrinos which may be witnesses of the most
violent phenomena in the Universe. With their very small probability of
 interactions and their neutral nature, they can come unaffected from the
 most distant parts of the Universe. 
Theoretical predictions give two regions of interest: from $10^{14}$ eV to $10^{17}$ eV for the contribution of AGN's, and from $10^{18}$ eV to $10^{20}$ eV for GZK neutrinos, the ones created but the high energy cosmic rays interacting with the CMB.

 Moreover, the study of high energy neutrinos may be used, not only to find 
  sources, but also to constrain some properties
 of these elusive particles.
For example, limits on oscillations can be achieved for ${\delta}m^{2}$ unreachable otherwise, and one can dream of searching for cosmic neutrinos through the Z$^{0}$ resonant absorption.

\section{Charged cosmic rays}

The spectrum of cosmic rays as it is measured to-day, extends 
over more than 30 orders of magnitude in flux and reaches energies of macroscopic 
size.

  The highest energy events are at the origin of a puzzle. 
Protons above $5.10^{19}$ eV undergo the GZK (Greisen-Zatsepin-Kuzmin) effect, namely their interaction
cross-section with the CMB background is such that one interaction length
amounts to about 50 Mpc. This is only 1\% of the size of the Universe and there are no
known sources energetic enough in this close neighbourhood. The GZK effect should
show as a pile-up of events followed by a cut-off at the critical energy.
Experimentally, a few events are seen above the cut-off, and there is no unique explanation for their 
existence. In 'top-down' models, these events are interpreted as traces of topological
defects or very high mass remnants of the Big Bang.

In order to shed some light on the problem, it is essential to increase 
the present statistics. Here comes the difficulty: the flux is 
extremely low, on the order of 1 cosmic per km$^{2}$ and century.

The Auger observatory is instrumenting 3000 km$^{2}$ in the Argentinian pampa, thus
hoping to collect some 30 events per year in the interesting energy region. The set-up should be operational in a couple of years.

Euso projects to use an optical telescope on the International Space 
Station. From 380 km altitude, it is able to watch 150000 km$^{2}$ of 
atmosphere and should detect some 800 events par year through 
the fluorescence light emitted along the showers generated by the passage
of the cosmic rays.

\section{The problems of neutrino detection}

Very ambitious projects also exist for neutrino detection. The aim is to instrument large 
volumes of up to 1 km$^{3}$ under the ice of the South Pole or deep underwater.
 However this technique is only efficient for the detection of up-going 
muons produced in the interactions of $\nu_{\mu}$. The energy resolution
on the detected muon is based on catastrophic bremsstrahlung and is limited
typically to a factor of 5.

The difficulty of detecting neutrinos comes from their very  
small cross-section necessitating large targets. Even at an energy of $10^{20}$ eV, the cross-section reaches
only $10^{-31}$ cm$^{2}$. The probability of interacting while traversing the atmosphere
is only 10$^{-4}$ . The most optimistic predictions of neutrino fluxes give no hope of seeing events coming down from the sky.
 Thus it is necessary to use a denser target. The earth is an obvious 
candidate but here comes another problem. At very high energies the earth becomes  
opaque to neutrinos. The cut-off happens at 100 TeV. Above this energy, $\nu_{e}$  and  $\nu_{\mu}$ 
are absorbed if they cross the earth through its center. The case of the  $\nu_{\tau}$ 
is more favorable because of the multibang phenomenon. An effect similar
to the GZK one takes place: very high energy  $\nu_{\tau}$  can cross the earth but 
their energy is degraded and they all exit with a few $10^{14}$ eV. Of course,
  $\nu_{\tau}$ detection is only possible if neutrinos oscillate with maximum mixing since
only   $\nu_{e}$  and  $\nu_{\mu}$   are directly produced in known mechanisms, at least in first order.

The Auger observatory will be able to detect a few  $\nu_{\tau}$  events, by catching some of the products of the interaction
in the mountain close to the apparatus. For Euso the possibility of 
detecting neutrino interactions by up-coming showers in the atmosphere 
is hampered by the threshold which is too close to the cut-off
coming from earth opacity.

\section{A novel way to look at neutrinos}

In order to alleviate these constrains, it is proposed to search for neutrino interactions in a mountain. This allows a detector closer to the shower and a target thin enough. The leptons generated in the process have a chance to come out of the rock and they subsequently
 develop a shower through the atmosphere of a valley. The shower is 
detected by its Cerenkov emission in an optical device.
Cerenkov emission has the advantage over fluorescence to give collimated light. For example the Euso threshold is a few $10^{19}$ eV for fluorescence, it is only 
a few $10^{14}$ eV for Cerenkov light. But the subtended 
solid angle in a given set-up reduces to 10$^{-3}$ srd, because of collimation.

The telescope copies the one of Euso. Its optics covers a large field of view of 60 degrees. It could in principle be a 1 m$^{2}$ prototype of the final device (Eusino?). The detector must be fast, it is based on multianode photomultipliers. The background fixes the needed number of pixels. The night-glow is
measured equal to 200 photoelectrons per m$^{2}$, ns and srd. In order to get 
a 10\% random occupancy with a 10 ns gate, this requires about 20000 pixels (10\% of the full Euso).
 With a valley 20 km wide, each pixel sees a spot of less than 100 m $\times$ 100 m of the opposite face.

 Two such telescopes separated by a few 100 m and observing the same area would allow coincidences and stereoscopic information by TOF, thus reducing drastically the background.

\section{Expectations}

There are two ways to look for Cerenkov light: direct observation when the
detector lies in the illuminated cone, or reflected light.
The first method is adapted to point-like sources, the second to a diffuse
flux.

For the direct detection of Cerenkov light, one considers neutrinos traversing the moutain ridge in front of the telescope.
Cerenkov intensity for a given electromagnetic shower in the atmosphere
has been well studied by the high energy photon telescopes (Cat, Whipple...).
A shower of 300 GeV initiated about 10 km above earth generates 2 photoelectrons per m$^{2}$ over a full area of half a km$^{2}$. This gives a total of a few 10$^{5}$ photoelectrons. Higher energy showers give a linearly increasing flux of photons.
The threshold is dictated by the number of detected photons. If one requires a minimum of 10 photoelectrons over a few adjacent pixels, the energy threshold is  2.10$^{13}$ eV. At such energy the atmospheric flux starts to die down in front of the predicted AGN contributions. This method is well adapted to search for a neutrino emission coming from our galactic center or from AGN's like the Markarians.

For the indirect Cerenkov light, the idea is to take advantage of the flash of light reflected on the mountain facing the telescope. The albedo is quite large, specially on a snowy surface.
The reflection spreads light over 2$\pi$ srd. The set-up is now sensitive to large solid angles, but the telescopes catch a small fraction, 10$^{-10}$, of the emitted photons due to reflection. This pushes the threshold to higher values. To collect 10 photoelectrons in a telescope, the initial shower energy must be above 10$^{17}$ eV. This energy range is well adapted to GZK neutrinos.

\section{Conclusion}

The relatively modest experiment discussed here can explore an effective area of 50 km$^{2}$ due to the large field of view of the optical device. The effective target depends on the nature of the considered neutrino.

For $\nu_{\mu}$, the produced ${\mu}$ can cross large distances in the rock before it exits, but the Cerenkov light emitted along its trajectory in the atmosphere may not be intense enough.

For $\nu_{\tau}$ the situation is the most favourable. The ${\tau}$ can cross several km in the rock, then comes out of the mountain and decays over the valley giving a high energy shower. The effective target is huge, up to 500 km$^{3}$srd at high enough energies. With the technique of the reflected Cerenkov flash, and with the expectations of GZK neutrinos, a calculation gives some 10 events per year.

For  $\nu_{e}$, one can take advantage of the Landau-Pomeranchouk-Migdal effect which suppresses the development of electromagnetic showers in dense materials. At energies of 10$^{20}$ eV, an electron produced by the interaction of a  $\nu_{e}$ can cross up to 100 m of rock. When it reaches the air the shower develops immediately. Thus the effective target has a depth of 100 m.

The present idea is essentially adapted to the detection of $\nu_{e}$ and $\nu_{\tau}$ interactions. This is complementary to the physics reach of volume detectors like Amanda and Antares which are limited to $\nu_{\mu}$ interactions. The angular resolution is quite good reaching 2 mrad. Moreover, this technique allows a decent energy measurement, typically 30$\%$, and, if statistics are large enough, spectra can be measured.

\section{References}

For more reading:

The Auger observatory, www.auger.org/admin/

The Euso project, www.ifcai.pa.cnr.it/$\sim$EUSO/

R.J.Protheroe, astro-ph/9809144

T.J.Weiler, hep-ph/0103023

A.Letessier-Selvon, astro-ph/0009416

X.Bertou et al., astro-ph/0104452

D.Fargion, astro-ph/0101565

S.Iyer Dutta et al., hep-ph/0012350
\setcounter{section}{0}
\setcounter{subsection}{0}

\end{document}